\documentclass[fleqn,12pt,a4paper]{article}
\usepackage{geometry}
\usepackage{epsfig}
\begin{document}                
\title{ Wavefront and ray-density plots using seventh-order matrices }
\author{Jos\'e B. Almeida\\
\emph{Universidade do Minho, Escola de Ci\^encias}\\ \emph{
4700-320 Braga, Portugal}}
\date{}

%
\maketitle
\begin{abstract}                
The optimization of an optical system benefits greatly from a
study of its aberrations and an identification of each of its
elements' contribution to the overall aberration figures. The
matrix formalism developed by one of the authors was the object of
a previous paper and allows the expression of image-space
coordinates as high-order polynomials of object-space coordinates.
In this paper we approach the question of aberrations, both
through the evaluation of the wavefront evolution along the system
and its departure from the ideal spherical shape and the use of
ray density plots. Using seventh-order matrix modeling, we can
calculate the optical path between any two points of a ray as it
travels along the optical system and we define the wavefront as
the locus of the points with any given optical path; the results
are presented on the form of traces of the wavefront on the
tangential plane, although the formalism would also permit sagital
plane plots. Ray density plots are obtained by actual derivation
of the seventh-order polynomials.
\end{abstract}
\section*{Keywords} Aberration, wavefront, ray density, matrix optics,
computer optics.

\section{Introduction}
In previous papers \cite{Almeida98, Almeida99} it was shown that
it is possible to determine coefficients for matrix modeling of
optical systems up to any desired order, computing power being the
only limiting factor. The same paper lists the calculated
seventh-order coefficients for systems comprising only spherical
surfaces.

The optical path length (henceforth designated \emph{opl}) of any
ray is the sum of the path length multiplied by the medium
refractive index, for all the media that compose the optical
system. The matrix modeling of the optical system is based on
translations between reference planes and orientation changes at
the surfaces separating two different media. In the following
paragraphs we will show that it is possible to evaluate the
optical path for all the translations incurred by any ray and add
them up to get an overall \emph{opl} between any two points on any
ray path.

If a known wavefront is used as origin for the evaluation of all
\emph{opl}s, then all subsequent wavefronts are loci of points
equidistant from the first wavefront in \emph{opl} terms. It is
then a question of preference the choice of method to display the
wavefront shape. The traces on the tangential and sagital planes
lead to simplified calculations and we will show examples of the
former. Every departure from a spherical wavefront is a
manifestation of aberrations. The common choice for reference
sphere is one that is centered on the paraxial image point and
contains the center of the exit pupil \cite{Welford91}.

Ray-density plots are also useful diagnosis tools because they are
similar to the actual images that the system will produce. We will
use the analytical expressions of image-space coordinates to
produce those plots.

\section{Optical system model}
If complex coordinates are used, an axis symmetric optical system
is modeled in the seventh-order by a product of $40\times40$
square matrices with real elements, each describing a particular
ray transformation. The elementary transformations can be
classified in four different classes:
\begin{itemize}
  \item Translation: A straight ray path.
  \item Surface refraction: Change in ray orientation governed by Snell's law.
  \item Forward offset: Ray path between the surface vertex plane and the
  surface.
  \item Reverse offset: Ray path from the surface back to the vertex plane, along the refracted ray
direction.
\end{itemize}

The ray itself is described by a 40-element vector comprising the
monomials of the complex position and orientation coordinates that
have non-zero coefficients. The product of all the elementary
transformation matrices yields the system matrix which must be
right-multiplied by the incident ray vector to result in the exit
ray vector.

The construction of elementary transformation matrices is
facilitated by the method described previously \cite{Almeida99,
Kondo96}. It can then be assumed that for any system comprising
only spherical surfaces all the necessary coefficients are known
and the system is perfectly described up to the seventh-order. All
the equations presented in the following paragraphs, relating
complex ray coordinates in the form of polynomials, were evaluated
by matrix multiplication using the software Mathematica. The size
of the matrices and the complexity of the expressions imposes some
care on the choice of elements to display; we will usually show
just the matrix element or the expression relevant for the
explanation under way.

In an aberration free optical system the wavefronts should have a
spherical shape throughout, or could eventually be flat in a
limiting case \cite{Welford91, Slyusarev84, Born80}. The departure
from a spherical wavefront shape is the manifestation of
aberrations. In well designed systems a wavefront may have become
become aspherical to be partially corrected further along the
system. The study of the distortions introduced on the wavefront
by each of the elementary ray transformations can greatly
elucidate about the performance of a particular system and provide
clues for an optimization procedure. Walther \cite{Walther96,
Walther96:2, Walther99} has performed such optimizations using
eikonals and computer algebra; in this paper we use matrix
formulation for the determination of wavefront shape at any point
along a complex system.

The method consists on evaluating the \emph{opl} of the rays as
they are subjected to the successive transformations and adding
them up until any desired position along the system is reached;
the result is the charateristic function $V(X,S,z,z')$, $X = x + i
y$ being the complex position coordinate, $S = s + i t$ the
complex orientation coordinate and $z$ and $z'$ the positions of
reference planes on object and image space, respectively
\cite{Born80, Goodman95}; $s$ and $t$ are the direction cosines
relative to axes $x$ and $y$, respectively.

Point objects are defined by a set of fixed coordinates $(x,y,z)$
and so the total \emph{opl} for rays originating on a point object
depends only on the ray orientation and image plane position,
$V(S,z')$. The locus of points with any given value of the
\emph{opl}, expressed by the equation $V(S,z')= constant$,
constitutes a wavefront \cite{Born80} whose shape can be plotted
or compared to a reference sphere. Before we start considering
each of the elementary transformations in turn we have to
establish that in cases where the incident beam is parallel we
will evaluate the \emph{opl} from an incident plane wavefront and
find the locus of points with constant \emph{opl} difference.

We will start by defining a generalized ray of complex coordinates
$(X, S)$; this ray is described by the 40-element monomials vector
$\mathbf{X}\&$, built according to the rules explained by Kondo
\cite{Kondo96} and Almeida \cite{Almeida99}. If the ray is
subjected to a transformation described by matrix $\mathbf{M}$,
then the output ray has coordinates $(X', S')$ and is represented
by the monomials vector $\mathbf{X'}\&$, such that:
\begin{equation}
  \mathbf{X'}\& = \mathbf{M X}\&.
\end{equation}

In the case of a translation the orientation coordinate does not
change and the \emph{opl} for that transformation is obviously
given by:
\begin{equation}
  l = \frac{n d} {(1-S S^*)^{1/2}} =  \frac{n d} {(1-S' S'^*)^{1/2}}~,
  \label{eq:path}
\end{equation}
with $n$ being the refractive index of the optical medium, $d$ the
distance traveled along the optical axis and the asterisk is used
to represent conjugate. The product of one complex number by its
conjugate is obviously one means of finding the square of its
modulus.

\begin{figure}[htb]
    \centerline{\psfig{file=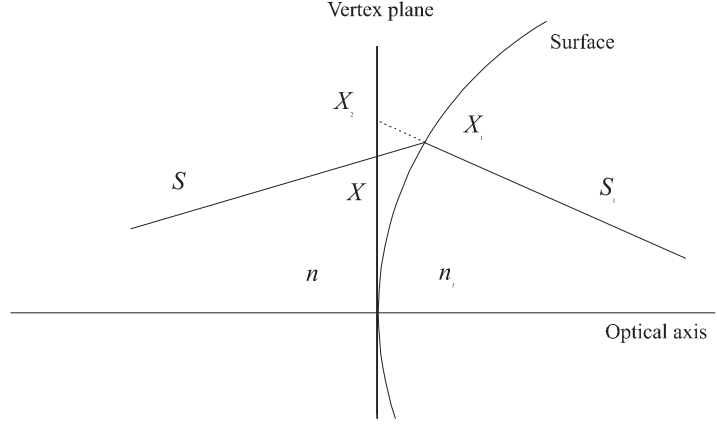, scale=0.8}}
\caption{\label{fig:offset} The ray intersects the surface at a
point $X_1$ which is different both from the point of intersection
of the incident ray with the plane of the vertex, $X$, and the point
of intersection of the refracted ray with the same plane, $X_2$. The
surface is responsible for three successive transformations: 1 -- an
offset from $X$ to $X_1$, 2 -- the refraction and 3 -- the offset
from $X_1$ to $X_2$.}
\end{figure}

A surface refraction introduces an orientation change but no path
length is involved and so it offers no contribution to the total
\emph{opl}. One optical surface contributes to the \emph{opl}
through both the forward and reverse offsets, which are not
conceptually different from the translation; both are translations
between the vertex plane and the surface, respectively in the
forward and the reverse directions, as represented in Fig.\
(\ref{fig:offset}). It is legitimate to use Eq.\ (\ref{eq:path})
to evaluate the path length contributions of these
transformations, as long as $d$ is not given a fixed value but is
evaluated for each incidence position; note, though, that there is
a refractive index change from the forward to the reverse offset,
besides the change in the ray orientation. In the following
section we will detail this procedure.

Plane waves with oblique orientation must be dealt with separately.
As the ray coordinates are referenced to planes normal to the
optical axis and there are phase differences between the plane wave
rays that intercept the reference plane at various points, those
phase differences must be accounted for by an \emph{opl} given by:
\begin{equation}
  l_0 = n (X X^* S S^*)^{1/2} = n \left|X\right| \left|S\right|~.
  \label{eq:pathpar}
\end{equation}
There is an implied assumption that the \emph{opl} is zero for the
ray that crosses the reference plane on the optical axis.

\section{Single refractive surface}
We first consider the case of a single surface with parallel
incidence. According to the previous argument, the first
\emph{opl} that has to be considered is $l_0$ given by Eq.\
(\ref{eq:pathpar}), which accounts for the phase differences of
the incident beam when it crosses the surface vertex plane; this
will obviously vanish if the rays are parallel to the optical
axis, which can always be verified by a single surface, if the
axis is chosen appropriately.

For the position coordinate of the ray after the forward offset we
refer to Fig.\ (\ref{fig:offset}) and use the coefficients given
by Almeida \cite{Almeida99}:
\begin{eqnarray}\label{eq:forward}
 X_1 &=&X + {\frac{S \left( 8 + 4 S S^* +
        3 {S^2} {{S^*}^2} \right)  X
      X^*}{16 r}} +
  {\frac{S \left( 2 + S S^* \right)  {X^2}
      {{X^*}^2}}{16 {r^3}}} +
  {\frac{S {X^3} {{X^*}^3}}{16 {r^5}}} \nonumber\\
  &=&X + \frac{S}{16} \left[\left(8 + 4 {\left|S\right|}^2 +
  3 {\left|S\right|}^4\right)\frac{{\left|X\right|}^2}{r} +
  \left(2 + {\left|S\right|}^2\right)\frac{{\left|X\right|}^4}{r^3} +
  \frac{{\left|X\right|}^6}{r^5} \right]~,
\end{eqnarray}
where $r$ represents the surface curvature radius.

In order to use Eq.\ (\ref{eq:path}) we must first find $d$ in
terms of the incidence point $X_1$; this is done by the following
equation:
\begin{equation}\label{eq:surf}
  d_1 = r - \left(r^2 - X_1 X_1^* \right)^{1/2} = r - \left(r^2 - {\left|X_1\right|}^2 \right)^{1/2}~.
\end{equation}
Now we can substitute Eq.\ (\ref{eq:surf}) in Eq.\ (\ref{eq:path})
to obtain the forward offset path length $l_1$.

After refraction the ray's orientation coordinate is changed according to
Snell's law; in the seventh-order approximation the new coordinate is given by:
\begin{eqnarray}\label{eq:direction}
 S_1 &=& \nu S + {\frac{\left( -1 + \nu \right)  X}{r}} +
  {\frac{\left( -\nu + {\nu^2} \right)  X S S^*}
    {2 r}} + {\frac{\left( -\nu + {\nu^2} \right)  {X^2}
      S^*}{2 {r^2}}}\nonumber \\
      && +  {\frac{\left( -\nu + {\nu^4} \right)  X {S^2}
      {{S^*}^2}}{8 r}} +
  {\frac{\left( -{\nu^2} + {\nu^4} \right)  {X^2} S
      {{S^*}^2}}{4 {r^2}}} +
  {\frac{\left( -{\nu^2} + {\nu^4} \right)  {X^3}
      {{S^*}^2}}{8 {r^3}}} \nonumber\\
      && +  {\frac{\nu \left( -1 + {\nu^5} \right)  X {S^3}
      {{S^*}^3}}{16 r}} +
  {\frac{{\nu^2} \left( -1 - 2 {\nu^2} + 3 {\nu^4} \right)  {X^2} {S^2}
      {{S^*}^3}}{16 {r^2}}} \nonumber\\
      && +  {\frac{3 {\nu^4} \left( -1 + {\nu^2} \right)  {X^3} S
      {{S^*}^3}}{16 {r^3}}} +
  {\frac{\left( -{\nu^4} + {\nu^6} \right)  {X^4}
      {{S^*}^3}}{16 {r^4}}} +
  {\frac{S \left( -\nu + {\nu^2} \right)  X X^*}
    {2 {r^2}}} \nonumber\\
      && + {\frac{\left( -\nu + {\nu^2} \right)  {X^2}
      X^*}{2 {r^3}}} +  {\frac{\left( -{\nu^2} + {\nu^4} \right)  X X^* {S^2}
      S^*}{4 {r^2}}} \nonumber\\
      && +
  {\frac{\nu \left( 1 - 3 \nu + 2 {\nu^3} \right)  {X^2} X^* S
      S^*}{4 {r^3}}} +  {\frac{\left( -{\nu^2} + {\nu^4} \right)  {X^3} X^*
      S^*}{4 {r^4}}} \nonumber\\
      && +
  {\frac{{\nu^2} \left( -1 - 2 {\nu^2} + 3 {\nu^4} \right)  X X^* {S^3}
      {{S^*}^2}}{16
      {r^2}}} + {\frac{\nu
      \left( 1 - 10 {\nu^3} + 9 {\nu^5} \right)  {X^2} X^* {S^2}
      {{S^*}^2}}{16
      {r^3}}} \nonumber\\
      && + {\frac{{\nu^2}
      \left( 2 - 11 {\nu^2} + 9 {\nu^4} \right)  {X^3} X^* S
      {{S^*}^2}}{16
      {r^4}}} + {\frac{3 {\nu^4} \left( -1 + {\nu^2} \right)  {X^4} X^*
      {{S^*}^2}}{16
      {r^5}}} \nonumber\\
      && + {\frac{\left( -{\nu^2} + {\nu^4} \right)  X
      {{X^*}^2} {S^2}} {8 {r^3}}} +
  {\frac{\left( -{\nu^2} + {\nu^4} \right)  {X^2}
      {X^*}^2 S} {4 {r^4}}} +  {\frac{\left( -\nu + {\nu^4} \right)  {X^3}
      {{X^*}^2}}{8 {r^5}}} \nonumber\\
      && +
  {\frac{3 {\nu^4} \left( -1 + {\nu^2} \right)  X {{X^*}^2} {S^3}
      S^*}{16
      {r^3}}} + {\frac{{\nu^2}
      \left( 2 - 11 {\nu^2} + 9 {\nu^4} \right)  {X^2} {{X^*}^2} {S^2}
      S^*}{16
      {r^4}}} \nonumber\\
      && + {\frac{\nu \left( 1 - 10 {\nu^3} + 9 {\nu^5} \right)
      {X^3} {{X^*}^2} S S^*}
      {16 {r^5}}} + {\frac{{\nu^2}
      \left( -1 - 2 {\nu^2} + 3 {\nu^4} \right)  {X^4} {{X^*}^2}
      S^*}{16
      {r^6}}} \nonumber\\
      && + {\frac{{\nu^4} \left( -1 + {\nu^2} \right)  X
      {{X^*}^3} {S^3}}{16 {r^4}}} +  {\frac{3 {\nu^4} \left( -1 + {\nu^2} \right)  {X^2}
      {{X^*}^3} {S^2}}{16 {r^5}}}\nonumber \\
      && +
  {\frac{{\nu^2} \left( -1 - 2 {\nu^2} + 3 {\nu^4} \right)  {X^3}
      {{X^*}^3} S}{16 {r^6}}}  +  {\frac{\left( -\nu + {\nu^6} \right)  {X^4}
      {{X^*}^3}}{16 {r^7}}}~,
\end{eqnarray}
where $\nu$ represents the refractive index ratio from the first medium to the
second.

The ray could now be traced back to the vertex plane along the
$S_1$ direction and the respective \emph{opl} calculated. We
prefer not to do this but rather to use an equivalent procedure
which consists on evaluating the next translation \emph{opl} from
the point $X_1$ on the surface instead of from the vertex plane.
The rays will now follow a straight path to the image plane at
some distance $z'$ from the surface; we calculate the
corresponding \emph{opl} ($l_2$) by means of Eq.\ (\ref{eq:path})
with $d$ replaced by $d_2=z' - d_1$ and $S$ replaced by $S_1$
taken from Eq.\ (\ref{eq:direction}).

We are now able to evaluate the total path length, $l_t$, in any
position in the second medium, just by adding the three
contributions $l_t = l_0 + l_1 + l_2$.

\section{Wavefront plots}
In the previous paragraph we established the method to evaluate
the path length for any ray as it intercepts any given reference
plane along the optical axis. In fact we defined a function of
$l_t(X,z')$ which is no other than the characteristic function
linking points on a wavefront in object space to points on a
reference plane in image space. In order to define the wavefront
surface we must specify a reference value for the characteristic
function and find the locus of the points where that reference
value holds; for convenience we take the value for the ray that
intercepts the reference plane on the optical axis and call this
$l_r$.

\begin{figure}[htb]
    \centerline{\psfig{file=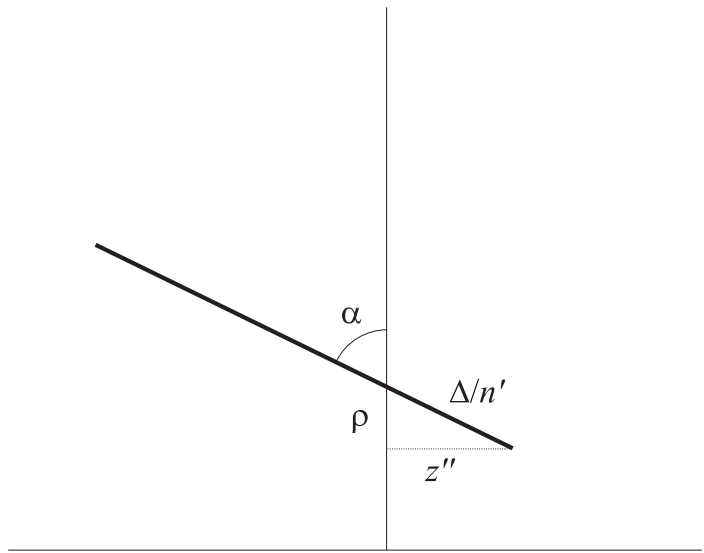, scale=0.8}}
\caption{\label{fig:delta} The figure represents a ray crossing a
reference plane normal to the optical axis; the plane of the figure
is not necessarily a meridional plane but it is rather the plane
containing the ray, which is normal to the reference plane.}
\end{figure}

Fig.\ \ref{fig:delta} represents a ray crossing a reference plane
normal to the optical axis; the plane of the figure is not
necessarily a meridional plane but it is rather the plane defined
by the ray and the normal to the reference plane on the point of
intersection. The ray coordinates on the point of intersection are
$(X', S')$ and the medium refractive index is $n'$; the optical
path difference is given by the difference $\Delta = l_r - l_t$.
If we were to follow along the ray the distance $\Delta / n'$ we
would find a point with the same \emph{opl} as the reference; this
point is necessarily on the same wavefront as the reference point.

From the figure we see that the projection of the distance $\Delta
/ n'$ on the reference plane is given by:
\begin{equation}\label{eq:rho}
  \rho =\frac{\Delta}{n'}\cos{\alpha}~.
\end{equation}
The factor $\cos{\alpha}$ can be decomposed on the direction
cosines relative to axes $x$ and $y$, leading to two components
$\rho_x$ and $\rho_y$, which must be added to the position
coordinates of the intersection point in order to obtain the
coordinates of the wavefront point; in complex notation it is:
\begin{equation}\label{eq:waveabs}
  X'' = X' + \frac{\Delta}{n'}S'~.
\end{equation}

The position of the wavefront point relative to the reference
plane is given by $z''$, according to the equation:
\begin{equation}
  z'' =\frac{ \Delta}{n'} \sin{\alpha}~;
\end{equation}
again in complex notation this can be rewritten:
\begin{equation}\label{eq:waveord}
  z'' =\frac{ \Delta}{n'} {\left(1 - S' {S'}^*\right)}^{1/2} =
  \frac{ \Delta}{n'} {\left(1 - {\left|S'\right|}^2\right)}^{1/2}~.
\end{equation}

The two equations (\ref{eq:waveabs}) and (\ref{eq:waveord}) define
a surface whose points have all the same optical path and so, by
definition, they are the wavefront equations.

\section{Numerical example}
For this example we chose a convex spherical surface of {1~m}
radius, which marks the boundary between air and a 1.5 refractive
index optical medium, upon which impinges a bundle of parallel
rays; the optical axis is chosen to be the line containing the
center of curvature which is parallel to the impinging rays. This
simple optical system has a paraxial focal distance of {3~m} and
the paraxial focus is the center of all the aberration--free
wavefronts considered after refraction.

We want to depict the wavefront shape through its trace on the
meridional plane; this allows an important simplification, as the
rays' position coordinate has null imaginary component and is thus
represented by the real component $x$; furthermore, the
orientation coordinate is zero because all the impinging rays are
parallel to the optical axis. As a result we have $x_1 = x$ and
from Eq.\ (\ref{eq:surf}):
\begin{equation}\label{eq:forward1}
  l_1 = d_1 = r - (r^2 - x_1^2)^{1/2}~.
\end{equation}

\begin{figure}[htb]
    \centerline{\psfig{file=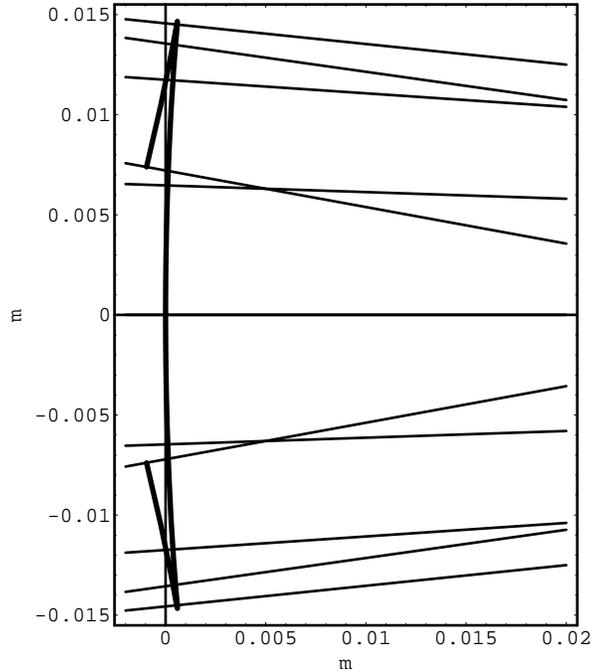, scale=0.8}}
\caption{\label{fig:surf} Meridional wavefront trace for a single
refracting surface, superimposed on the traces of meridional rays.
Notice that the ends of the wavefront are folded and show a convex
curvature, indicating spherical aberration.}
\end{figure}

The orientation coordinate after refraction, $S_1$, is real for
all rays on the meridional plane and so it is represented in lower
case: $S_1 = s_1$. This was evaluated by matrix multiplication but
we could just as well have used Eq.\ (\ref{eq:direction}) with
suitable substitutions. We applied Eq.\ (\ref{eq:path}) to
evaluate the optical path contribution of the translation from the
surface vertex plane to a reference plane located {2.8~m} after
the surface; the refractive index was set to $n = 1.5$ and the
distance was set to $d_2 = 2.8-d_1$. Eqs.\ (\ref{eq:waveabs}) and
(\ref{eq:waveord}), with real position coordinates, were used to
evaluate the curve of the wavefront trace which was then plotted
as shown in Fig.\ \ref{fig:surf} superimposed on the traces of
meridional rays; these are naturally normal to the wavefront in
every point. We notice that the ends of the wavefront are folded
and show a convex curvature, indicating spherical aberration. The
points on the curve with zero curvature radius are points on a
caustic arising from the crossing of rays with different
directions.

\section{Single lens}
We turn our attention now to a thin lens with oblique incidence.
The lens is convex on the first surface and flat on the second
surface, the convex surface has a curvature radius of {31.123~mm}
and the center thickness is {5.8~mm}; the glass is BK7, defined as
having a refractive index of 1.5168. This lens has a nominal focal
distance of {60~mm}. The rays incident upon the lens form a
parallel bundle with a direction cosine $s=0.1$.

The only added complication to the situation of the single surface
results from the consideration of the second surface, which marks
the transition from glass to air with no associated curvature. The
optical axis is now clearly identified by the line normal to the
flat surface and containing the first surface's center of
curvature and cannot be aligned with the direction of incidence.
Oblique incidence promotes the emergence of the various aberration
terms but does not imply any new equations.

\begin{figure}[htb]
    \centerline{\psfig{file=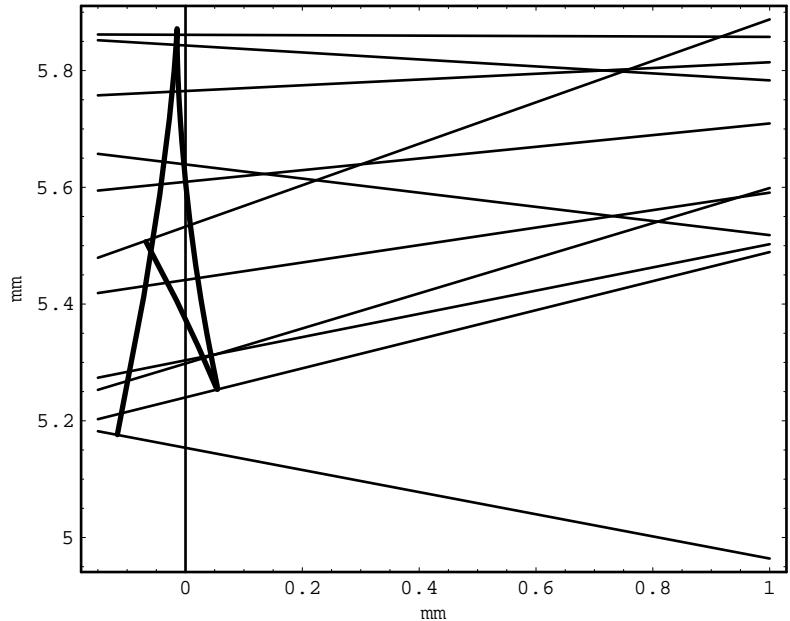, scale=0.8}}
\caption{\label{fig:lens} Meridional wavefront trace for a single
lens, superimposed on the traces of meridional rays. The sharper
bend of the upper end is an indication of coma.}
\end{figure}

The wavefront is studied at a distance of {52~mm} and its
meridional trace is plotted on Fig. \ref{fig:lens}. Again we
notice that the rays are normal to the wavefront and that the ends
of this are folded backwards. The sharper bend of the upper end is
an indication of coma. The other aberration terms are not clearly
noticeable on the figure because in the case of astigmatism we
would have to compare with the sagital plot and in the cases of
field curvature and distortion the wavefront is still spherical
but its center is shifted from the paraxial position.

\section{Ray-density plots}
In order to study the ray-density plots we use the lens of the
previous example wit the image plane moved to a position just past
the tangential focus, i.e. {56~mm}, because this is a natural
position and also because all the rays are divergent from this
position onwards. This avoids the complication of having to deal
with overlapping wavefront folds, each of them contributing
independently to the overall ray-density.

The overall system matrix for the lens above followed by a
straight path to the image plane was evaluated with the help of
Mathematica and then right-multiplied by the input ray vector
$\mathbf{X}\&$ composed with the variable complex position
coordinate, $X$, and fixed orientation coordinate, $s=0.1$. The
result is a 40--element vector, the first of which is a 7th--order
polynomial on $X$, representing the dependence of the point of
intersection on the image plane on the input variable; we call
this $X'$.

If it is established that the input beam has a uniform
ray-density, then the image-plane ray-density is given by:
\begin{equation}\label{eq:dens}
i = {1}/\left|{\frac{\mathrm{d}X'}{\mathrm{d}X}}\right|~.
\end{equation}
Now, $X$ being a complex coordinate we can express it in the
exponential form as $X = \chi e^{i \theta}$ and plug this into
Eq.\ (\ref{eq:dens}) to get:
\begin{equation}\label{eq:dens2}
i = {1}/\left|{\frac{\mathrm{d}X'}{\mathrm{d}\chi}-
\frac{i}{\chi}\frac{\mathrm{d}X'}{\mathrm{d}\theta}}\right|~.
\end{equation}

\begin{figure}[htb]
    \centerline{\psfig{file=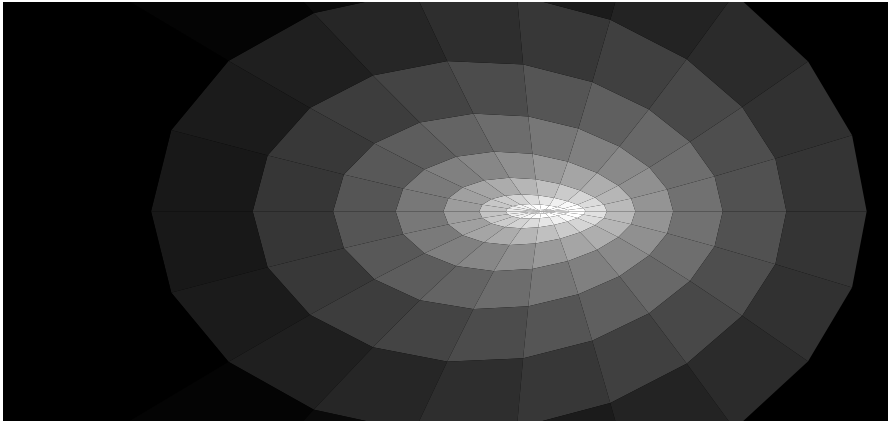, scale=0.8}}
\caption{\label{fig:density} Ray density plot on a plane near the
tangential focus.}
\end{figure}

The value of the ray-density given by Eq.\ (\ref{eq:dens2}) was
evaluated and plotted as shades of gray on a logaritmic scale on
the positions corresponding to the image coordinate $X'$, as shown
on Fig.\ \ref{fig:density}. The image is just as one would expect
from a lens focusing an oblique beam of light.

\section{Conclusion}
Previous results had shown that optical systems could be modeled
with matrices up to any desired order of approximation and the
necessary coefficients for axis-symmetrical systems built with
spherical surfaces had already bean reported. Those results have
now been used to evaluate aberrations in non-standard ways.

An implementation of the seventh-order matrix algorithm in
Mathematica allows the construction of algebraic models for very
complex systems, which can be used in various ways to judge their
performance and quality.

The possibility of plotting wavefront shapes at any point along a
complex optical system was demonstrated with two simple examples
but the same procedure could be used in more complex situations.
Ray-density plots were also demonstrated, these providing a
visualization of the actual image of point objects. It is expected
that ray-density plots can be integrated for extended objects,
thus yielding the expected aberrated images given by real optical
systems.
\nocite{Bass95} \pagebreak
  \bibliography{aberrations}   
  \bibliographystyle{OSA}   

\end{document}